\documentstyle[psfig]{mn}

\def \msol {\rm{M}$_\odot$}

\def \kms{km~$\rm{s}^{-1}$}
\def \cc{$\rm{cm}^{-3}$}

\title[Outflow collimation in YSOs]
{Outflow collimation in young stellar objects}

\author[Mellema \& Frank]
{Garrelt Mellema$^1$ and Adam Frank$^2$\\
$^1$Stockholm Observatory, S-13336 Saltsj{\"o}baden, Sweden;\\ 
    email: garrelt@astro.su.se\\
$^2$Department of Physics and Astronomy, Bausch and Lomb Building,
  University of Rochester, Rochester, NY 14627-0171, USA;\\ 
  email: afrank@alethea.pas.rochester.edu}
\date{Accepted 1997 July 7. Received 1997 June 20; in original form 
1996 December 31}

\pagerange{\pageref{firstpage}--\pageref{lastpage}}
\pubyear{1997}

\begin{document}

\maketitle
\label{firstpage}
\begin{abstract}
In this paper we explore the effect of radiative losses on purely
hydrodynamic jet collimation models applicable to Young Stellar Objects
(YSOs). In our models aspherical bubbles form from the interaction of a
central YSO wind with an aspherical circum-protostellar density distribution.
Building on a previous non-radiative study (Frank \& Mellema 1996) we
demonstrate that supersonic jets are a natural and robust consequence of
aspherical wind-blown bubble evolution. The simulations show that the
addition of radiative cooling makes the hydrodynamic collimation mechanisms
studied by Frank \& Mellema (1996) more effective.  We find a number of
time-dependent processes contributing to the collimation whose relative
strength depends on the age of the system and parameters characterising the
wind and the environment.  As predicted by Frank \& Mellema (1996) the
flow-focusing at an oblique inner shock becomes more effective when radiative
cooling is included. An unexpected result of this is the
production of cool ($T < 10^4$ K), dense ($n \approx 10^4$ cm$^{\rm -3}$)
jets forming through conical converging flows at the poles of the bubbles.
For steady winds the formation of these jets occurs early in the bubble
evolution.  At later times we find that the dynamical and cooling time scales
for the jet material become similar and the jet beam increases in temperature
($T \approx 10^6$ K).  The duration of the cool jet phase depends on the mass
loss rate, $\dot M_{\rm w}$, and velocity, $V_{\rm w}$, of the wind.  High
values of $\dot M_{\rm w}$ and low values of $V_{\rm w}$ produce longer cool
jet phases.

Since observations of YSO jets show considerable variability in the jet beam
we present a simple one-dimensional (1-D) model for the evolution
of a variable wind interacting with an accreting environment.  We
find that the accretion ram pressure can halt the expansion of the bubble on
time scales comparable to the periodicity of the wind and length scales less
than 100 AU, the approximate observed scale for YSO jet collimation.
These models indicate that, in the presence of a varying protostellar
wind, the hydrodynamic collimation processes studied in our simulations can
produce cool jets with sizes and time scales consistent with observations.
\end{abstract}


\section{Introduction} 
A large fraction of stars begin their lives in the midst of narrow supersonic
streams of gas or `jets'. These jets are a common phenomenon and are observed
to carry away large amounts of energy and momentum from the central regions
of Young Stellar Objects (YSOs).  The propagation of these jets has been well
studied both analytically \cite{RaKof92} and with sophisticated numerical
tools \cite{BlFrKo90,StoNor94a}. However, the processes responsible for
collimation remains an unsolved problem.  The current consensus favours
collimation through magnetohydrodynamical (MHD) effects
\cite{Kon89,Pud91,Shuea94}.  While these models are promising they may
require conditions which are not achieved in real YSO environments (such as
particular field geometries and long collimation scale lengths). In addition
some numerical simulations based on these MHD models show that while winds
can effectively be produced collimation into a steady jet may be more
difficult to achieve \cite{Romanovaea96}. Hydrodynamic collimation models
such as DeLaval nozzles \cite{Kon82,RaCan89} have fallen out of favour
because of length scale requirements \cite{KoRu93}, stability considerations
\cite{KoMc92}, and the lack of an energy source to drive the outflow.
Especially for low mass stars one can show that the outflows cannot be
generated by radiation pressure. However, if one separates the issues of wind
{\it acceleration} and jet {\it collimation} this last point is not a real
objection. One can then have initial acceleration through an MHD process and
collimation through a hydrodynamic process. In weighing the relative
advantages of hydrodynamic and MHD collimation models it is clear that the
hydrodynamic models offer the advantage of being simpler in terms of
underlying physical processes and requirements on initial conditions.

Until recently the majority of jet collimation models, both hydrodynamic and
MHD, have been analytic. To reduce the complexity of the equations involved,
these models need simplifying assumptions which typically involve ignoring
the time-dependence in the equations, in other words steady state solutions.
This impliesignoring the dynamical feedback of the global flow
\cite{RaCan89} and freezing initial configurations such as exterior pressure
distributions.  These assumptions, though necessary to make initial progress,
are suspect from first principles. In addition, it has become clear that the
flows in YSO jets are {\it essentially} unsteady.  Recent observations of HH
jets show strong evidence for velocity variations within the jet beam
\cite{Hartea93}, \cite{Hartet96}.  Molecular outflows, which may be driven by jets
\cite{Chernea94}, also show evidence for episodic variations in outflow speed
and density \cite{BaTaCe94}. These observations suggest that the driving of
the YSO outflow and its interaction with the circumstellar environment are
fundamentally time-dependent processes, and time-dependent numerical models
are needed.

In a recent work Frank \& Mellema~\shortcite{FrMel96} (hereafter FM96)
revisited the issue of purely hydrodynamic collimation mechanism using high
resolution numerical simulations.  FM96 focused on the interaction of a wind
from a protostar wind with an aspherical (toroidal) circum-protostellar
environment.  Studies of planetary nebulae have demonstrated that this kind
of interaction can produce well collimated jets \cite{Ickeea92}.  The
mechanism has been called ``Shock-Focused Inertial Confinement'' (SFIC).  In
the SFIC mechanism it is the inertia of a toroidal environment rather than
its thermal pressure, which produces a bipolar {\it wind-blown bubble}.  The
bubble's reverse shock, which decelerates the wind, takes on an aspherical,
prolate geometry \cite{Eic82}.  The radially streaming central wind strikes
this prolate shock obliquely focusing it towards the polar axis and
initiating the jet collimation.  Other effects such as instabilities along
the walls of the bubble, help to maintain the collimation of the shocked wind
flow.

In an initial study Frank \& Noriega-Crespo~\shortcite{FrNor94} used
non-radiative numerical simulations to demonstrate that the SFIC mechanism
can produce jets in the context of YSOs. Peter \& Eichler~\shortcite{PeEi95}
studied the inertial confinement of fully formed jets in a more general
context.  FM96 carried out a more extensive study of the SFIC mechanism in
YSO systems.  The high resolution simulations presented in FM96 showed that a
central wind interacting with a toroidal shaped density distribution
naturally leads to the development of strongly collimated supersonic flows.
Toroidal density distributions are theoretically expected to form from the
collapse of a rotating cloud \cite{TSC84} or of a flattened filament
\cite{Hartmea96}, or of a magnetised cloud \cite{LiShu96}. In addition there
is also some observational evidence for such structures
\cite{LucRoch97,Kraemerea97}.

The collimated supersonic flows found in the simulations are accompanied by
all the usual features expected for gaseous jets: bow and jet shocks;
turbulent cocoons; crossing shocks and internal Mach disks.  Because the aim
of FM96 was to explore the basic physics of the SFIC mechanism in detail they
used simulations without cooling and then applied analytical models to
explicate the underlying dynamics.  In this way FM96 showed the dual nature
of the collimated flow as both a supersonic jet and a wind-blown bubble. More
importantly they also concluded that even a small degree of asphericity in
the reverse shock is sufficient to produce strong flow focusing
(cf.~Icke~1988). Using an analytical approximation to estimate the effects of
radiative cooling they found that with radiative cooling included these
shocks are capable of achieving collimation on length scales smaller than
100~AU, consistent with the {\it HST}\/ observations
\cite{Heathea96,Burea96}. Their analytical models also showed that flow
focusing becomes even more effective in cooling shocks (cf.~Icke~1988). This
allows the wind to be redirected into a jet without becoming subsonic.

While the results of FM96 were promising many questions remain to be
answered.  These include: the dynamical role of cooling; the effect of more
realistic environments; the connection with YSO observables.  Each of these
issues deserves considerable attention.  Our philosophy in pursuing this line
of research is to isolate domains of interest and use simulations as
numerical experiments to reveal and then articulate the underlying physical
processes. Following this strategy we focus here on the first question: the
dynamical effect of radiative cooling.  As we will demonstrate the addition
of cooling to the simulations produces dramatic changes in the flow which
{\it enhance} the previously studied non-radiative SFIC collimation process,
as was predicted in FM96.  What was not predicted is a new collimation
process, which operates when radiative cooling is included. This process,
first explored in a series of papers by Tenorio-Tagle, Cant{\'o} \&
R{\'o}\.zyczka~\shortcite{TenCanRoz88}, may be applicable to jet collimation
not only in YSOs but also in other objects, such as planetary nebulae
\cite{FrBaLi96,Mell96}.  Some of the results seen in our simulations are
found in a more general and more abstracted series of simulations done by
Peter \& Eichler~\shortcite{PeEi96}.  Their results confirm the efficacy of
the radiative collimation processes explored in a more dynamical context
here.

The organization of the paper is as follows: In Section 2 we describe the
numerical method and initial conditions used in our simulations.  In Section
3 we present and discuss the results of our simulations. In Section 4 we
address the wind variability.  Finally in Section 5 we present our
conclusions along with a discussion of some issues raised by the
simulations.

%
\section{Computational Methods and Initial Conditions}
The hydrodynamic interactions we wish to study are governed by the Euler
equations with a `sink' term due in the energy equation due to radiative
losses.
\begin{equation}
{\partial\rho \over \partial t} + \bmath{\nabla} \cdot \rho \bmath{ u} = 0\,,
\label{masscon}
\end{equation}
\begin{equation}
{\partial\rho \bmath{ u} \over \partial t} + \bmath{ \nabla} \cdot
\rho \bmath{ uu} = \bmath{\nabla}P\,,
\label{momcon}
\end{equation}
\begin{equation}
{\partial E \over \partial t} + \bmath{ \nabla} \cdot \bmath{ u} (E + p) = -
{x^2_{\rm i} \rho \over \bar m}^2 \Lambda(T) - 
{\partial x_{\rm i} \over \partial t}I_{\rm H}\,,
\label{encon}
\end{equation}
where
\begin{equation}
E = {1 \over 2} \rho \vert \bmath{ u} \vert^2 + {p \over (\gamma - 1)}\,,
\label{endef}
\end{equation}
\noindent and
\begin{equation}
p = {\rho k T\over \bar m} \,. 
\label{pdef}
\end{equation}
\noindent In the above equations ${\bar m}$ is the mean mass per particle,
$x_{\rm i}$ is the fraction of ionized hydrogen, and we take $\gamma = 5/3$.

\subsection{Numerical methods}
We have carried out our study using two different numerical codes each of
which is cast in a different coordinate geometry. All the simulations are
run in two dimensions, assuming cylindrical symmetry. The first code is
based on a Roe-solver method \cite{Melal91} and uses spherical coordinates
($R,\theta$).  The second code is based on the Total Variation Diminishing
(TVD) method of Harten~\shortcite{harten83} as implemented by Rye et al.~\shortcite{Ryuet95}.  The TVD
code solves the Euler equations in cylindrical coordinates ($r,z$).
Both codes are explicit methods for solving hyperbolic systems of
equations. Second order accuracy is achieved by finding approximate
solutions to the Riemann problem at grid boundaries.  Oscillations are
prevented by using a lower order monotone scheme near steep changes.

Previous experience \cite{FrMel94} has shown us the value of using two
different codes, based on different numerical methods and applied with
different geometries, to work on the same problem.  One of the most difficult
aspects of numerical studies is knowing to which extend one can trust the
results. Applying two codes to the same problem can quickly and convincingly
root out numerical artifacts.  We note that the TVD code in its present
configuration is more diffusive than the Roe solver code and this fact must
be considered in comparing their results in detail. The main point we wish to
make in comparing the two codes is that both produce hydrodynamically
collimated jets via the same mechanisms.

In both codes cooling is included using operator splitting and is
calculated from look-up tables for $\Lambda(T)$ taken from the coronal
cooling curve of Dalgarno \& McCray~\shortcite{DalMc72}. The treatment of the
cooling term is different in both codes. For the Roe solver the collisional
ionization equation for hydrogen is solved first and the computed value of
$x_{\rm i}$ is fed into equation \ref{encon}. The cooling source term is then
integrated into the solution through iteration. Cooling through collisional
ionization of hydrogen (the ${\partial x_{\rm i}\over\partial t}I_{\rm H}$
term) is also taken into account. Because of the iteration it is not
necessary to set a cooling time limit on the time step, although in practise
it helps to avoid numerical problems.

In the TVD code full ionization is assumed and the cooling is applied via an
integration of the thermal energy ($E_{\rm t}$) equation
\begin{equation}
{d E_{\rm t}\over dt} = - {\dot E_{\rm t}} = 
\left({\rho \over \bar m}\right)^2 \Lambda(T)\,.
\label{ethdot}
\end{equation}
For application to the simulations the solution to this equation takes the
form
\begin{equation}
{E_{\rm t}}^{n+1} = {E_{\rm t}}^{n}\exp(-{{\dot E_{\rm t}} \over {E_{\rm
t}}^{n}}\Delta t)\,,
\label{etsol}
\end{equation}
where the superscript $n$ refers to the time index ($t^n = n \Delta t$). The
term in the exponential can be expressed as as $\Delta t / \Delta t_{\rm c}$
where $\Delta t_{\rm c}$ is the local cooling time scale of the gas: $\Delta
t_{\rm c} = {\dot E_{\rm t}} / E_{\rm t}$.  This formulation has the
advantage protecting ${E_{\rm t}}^{n+1}$ from becoming negative in regions
of strong cooling.  In practice one must choose the time step $\Delta t$ so
as not to be in conflict with short cooling time scales.  We use
\begin{equation}
\Delta t = \min\lbrack\Delta t_{\rm hydro},1.5 \Delta t_{\rm c}\rbrack
\,,
\label{timestep}
\end{equation} 
where $\Delta t_{\rm hydro}$ is the hydrodynamic time step set by the
Courant condition.

Both codes have been tested against analytical models of wind-blown bubbles
\cite{KoMc92} and were found to recover appropriate values of various
shock positions and velocities.

\subsection{Initial conditions}
The initial conditions used in our simulations are identical to those used
in FM96 and we refer the reader to that paper for the details.  Our
simulations begin with a stationary isothermal environment characterized by
a aspherical (toroidal) density distribution.  We have used a distribution
$\rho(r,\theta)$ which creates a pseudo-disk of FWHM $= 90^\circ$.  The
input parameters for the environment are the mass of the central
protostellar object (taken to be $M = 1$\msol), the accretion rate ${\dot
M_{\rm a}}$ which determines the density in the environment, and finally,
the equator to pole density contrast $q = \rho(0^\circ)/\rho(90^\circ)$.  In
the FM96 it was found that supersonic jets were produced for $q > 7$.

As in FM96 we do not include infall velocity, nor the effects of gravity or
rotation.  As was noted in the introduction the aim is to isolate the effects
of radiative cooling on the collimation models. In Section 4 we present
analytical models that will address the effects of both gravity and accretion
ram pressure on the the bubble dynamics and jet collimation.  In addition the
next paper in this series will focus on the SFIC collimation in more
realistic environments \cite{Yorkea93,Hartmea96}.  

An even further complication would be a non-axisymmetric accreting
environment. How such an environment affects the flow cannot be addressed
without 3-D models which is beyond the scope of the present work. One may
also worry about the longevity of the toroidal density distributions.
However, both Li \& Shu~\shortcite{LiShu96} and Matsumoto, Hanawa \&
Nakamura~\shortcite{Mat97} find self-similar (i.e.\ scale free) toroidal
structures in their models, implying that the accretion flow will maintain
this geometry over a long time span.

The central protostellar wind is fixed in an inner sphere of grid cells. The
relevant input parameters are simply the mass loss rate ${\dot M_{\rm w}}$
and velocity $V_{\rm w}$ in the wind.

In Table 1 we list the relevant input parameters for the 5 simulations
presented in this paper.  We have carried out more than 30 simulations
including a sequence of simulations at different resolutions ($64\times 320$;
$128\times 640$; $256\times 1280$).  These convergence tests demonstrate that
the main collimation features are adequately resolved (though we have not
reached full convergence). A comparison between the $128\times 640$ and
$256\times 1280$ results shows that the higher resolution simulations reveal
more detail but do not show changes in overall flow pattern.

However, our flow solutions do contain strong cooling regions which often our
grids cannot adequately resolve. One must therefore be careful in
interpreting the results as under-resolved cooling zones may lead to grid
mixing. Problems like these can only be addressed by continued modelling at
higher resolution when additional computational resources become available.
However, seeing similar flow patterns in two different codes configured in
different geometries strengthens the argument that we are seeing real
physical effects rather than numerical artifacts. In fact very similar
results have also been found when using a PPM (Piecewise Parabolic Method)
code using an expanding grid to study strongly cooling wind-blown bubbles in
the context of planetary nebulae (Dwarkadas, private communication). Thus the
development of hydrodynamically collimated jets in wind-blown bubbles has
been found in two different sets of numerical experiments with three
different kinds of numerical tools.

\begin {table}
\caption {Initial Conditions For Runs A -- E}
\label{initcond}
\begin {tabular} {lllllll} 
{run}  & {${\dot M}_{\rm w}$} & {$V_{\rm w}$} & {${\dot M}_{\rm a}$} 
& {$q$} & {resolution} \\
A  &  $1 \times 10^{-7}$ & 350  & $1 \times 10^{-6}$ & 70 & 
$256 \times 1536$ \\

B  &  $2 \times 10^{-7}$ & 250 & $1 \times 10^{-5}$ & 70 & 
$256 \times 1280$ \\

C  &  $1 \times 10^{-7}$ & 450 & $5 \times 10^{-6}$ &  60 & 
$256 \times 1280$ \\

D  &  $1 \times 10^{-6}$ & 250 & $1 \times 10^{-7}$ &  50 & 
$250 \times 250$ \\

E  &  $1 \times 10^{-8}$ & 250 & $1 \times 10^{-9}$ &  50 & 
$250 \times 250$ \\
%

\end {tabular}
\end {table}

%
\section{Simulation Results}
Here we present the results of several numerical simulations, calculated
with both methods introduced above (TVD and Roe solver). We explore two
sequences in parameter space, runs A to C form a sequence in outflow velocity
(from 250 to 450~\kms), calculated with the TVD-method. Runs D and E
(calculated with the Roe solver) form a sequence in density, which serves to
illustrate the importance of cooling. See Table 1. We mainly concentrate on
describing the results of run A, and then point out some of the differences
with the other runs.

\begin{figure*}
\centerline{\psfig{figure=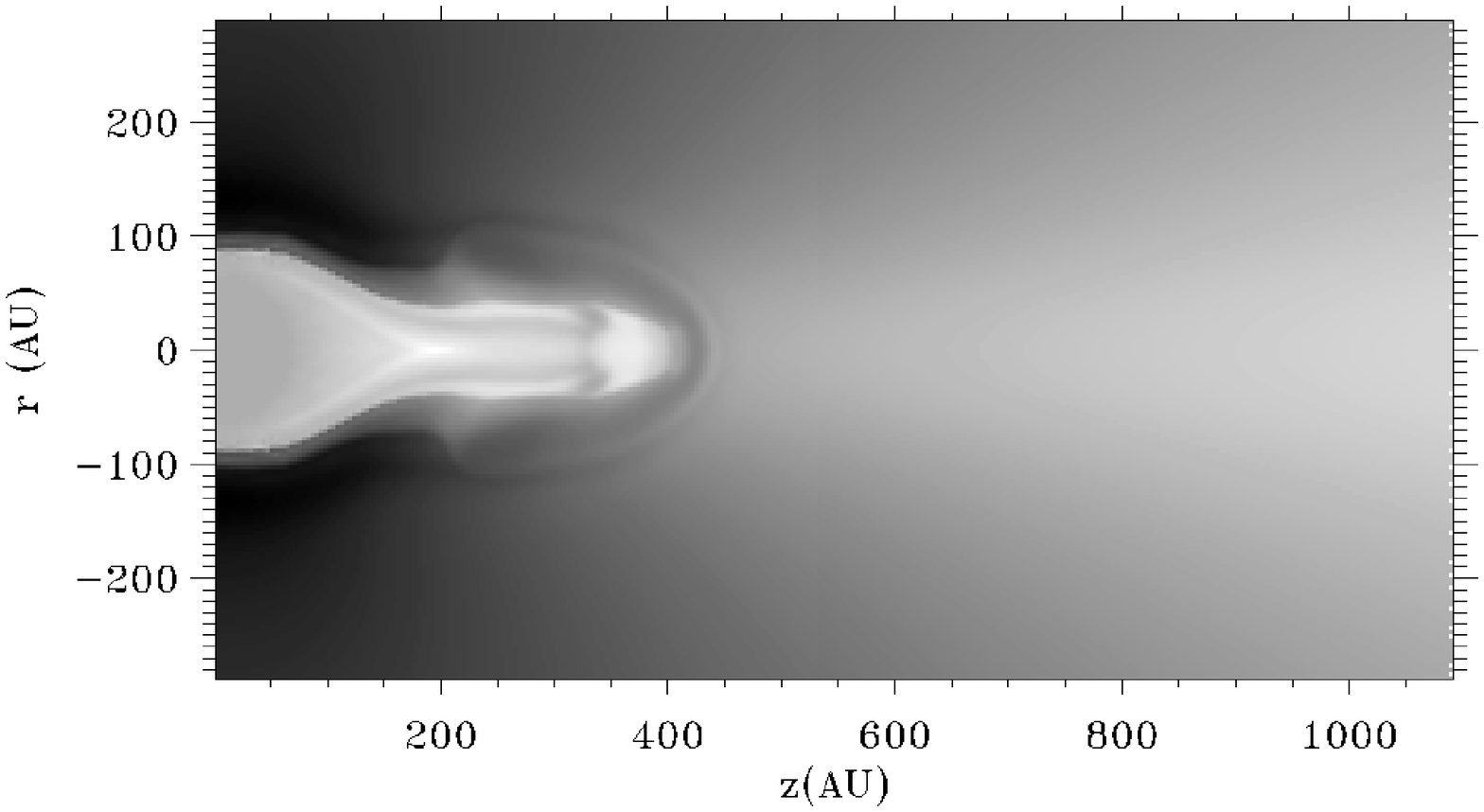,width=9cm}\psfig{figure=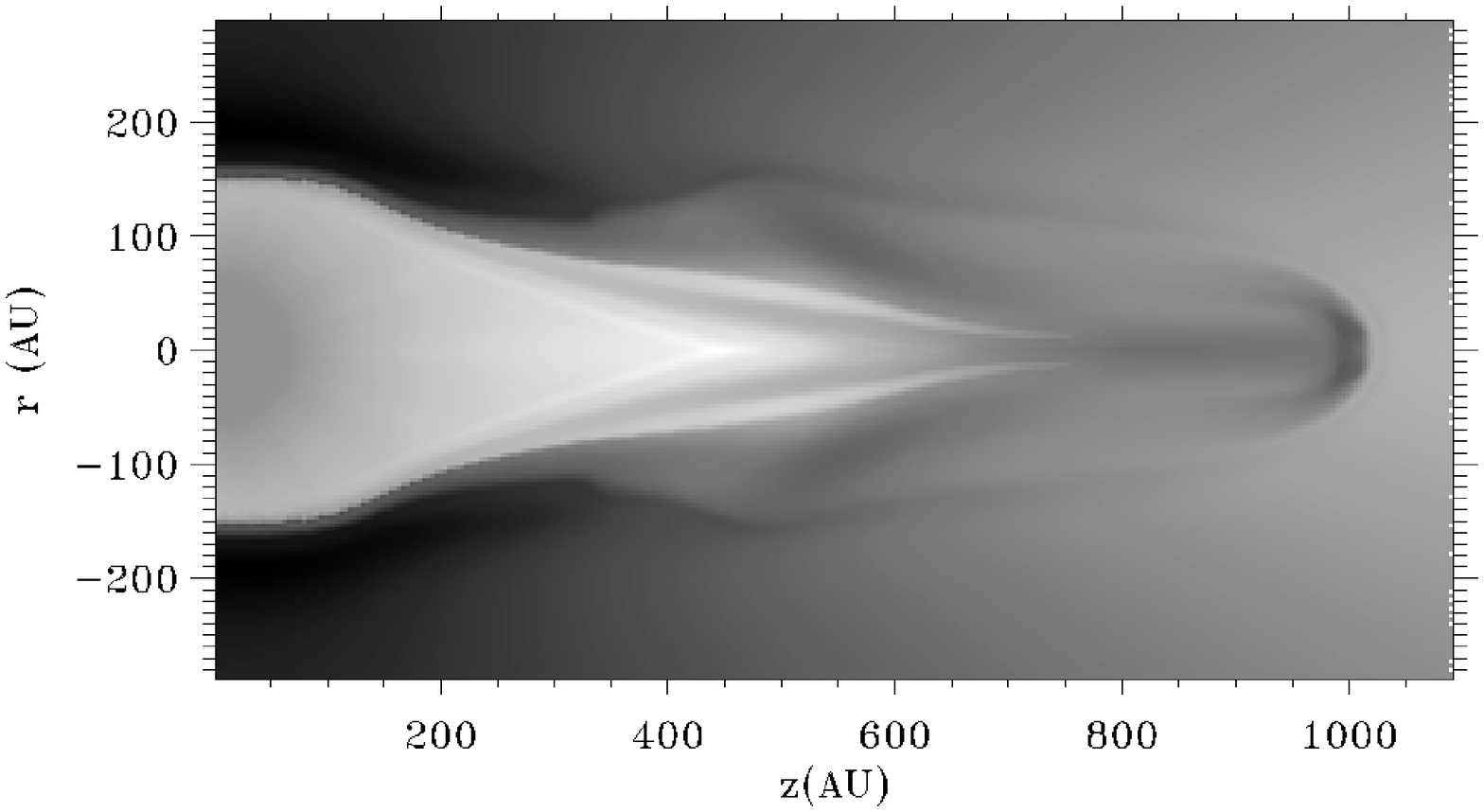,width=9cm}}
\centerline{\psfig{figure=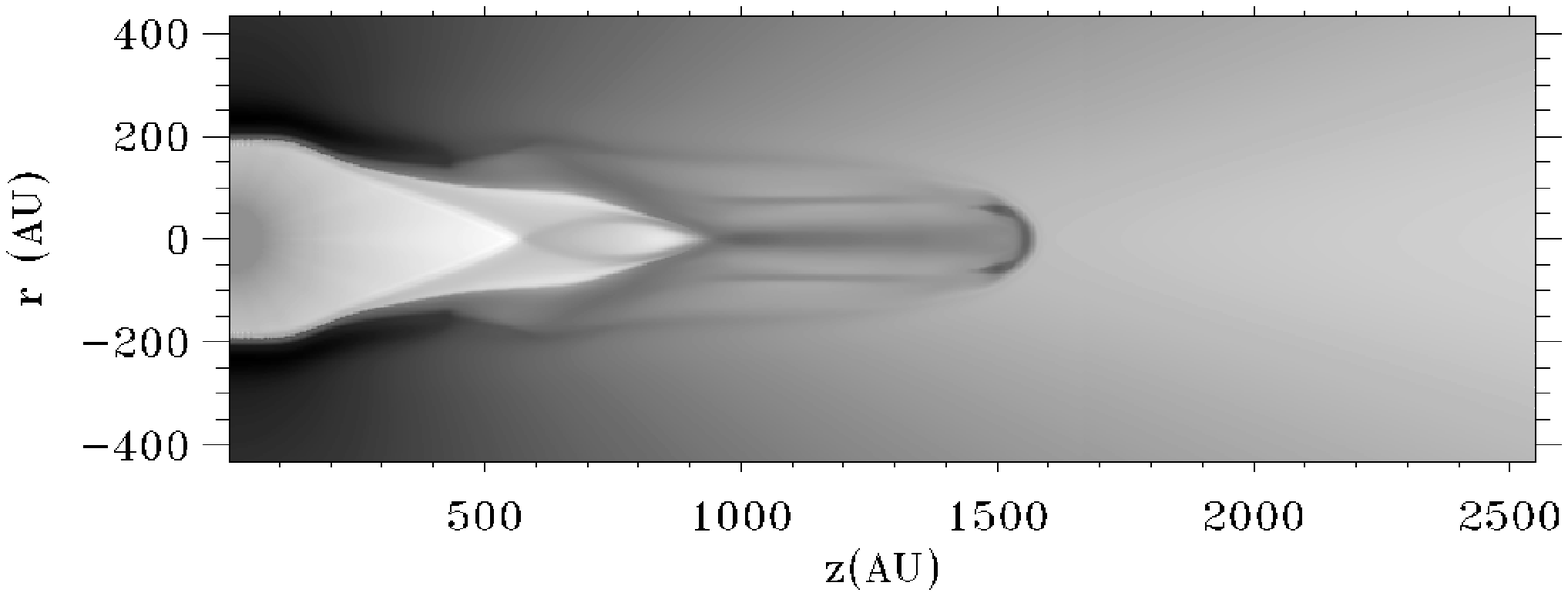,width=17cm}}
\centerline{\psfig{figure=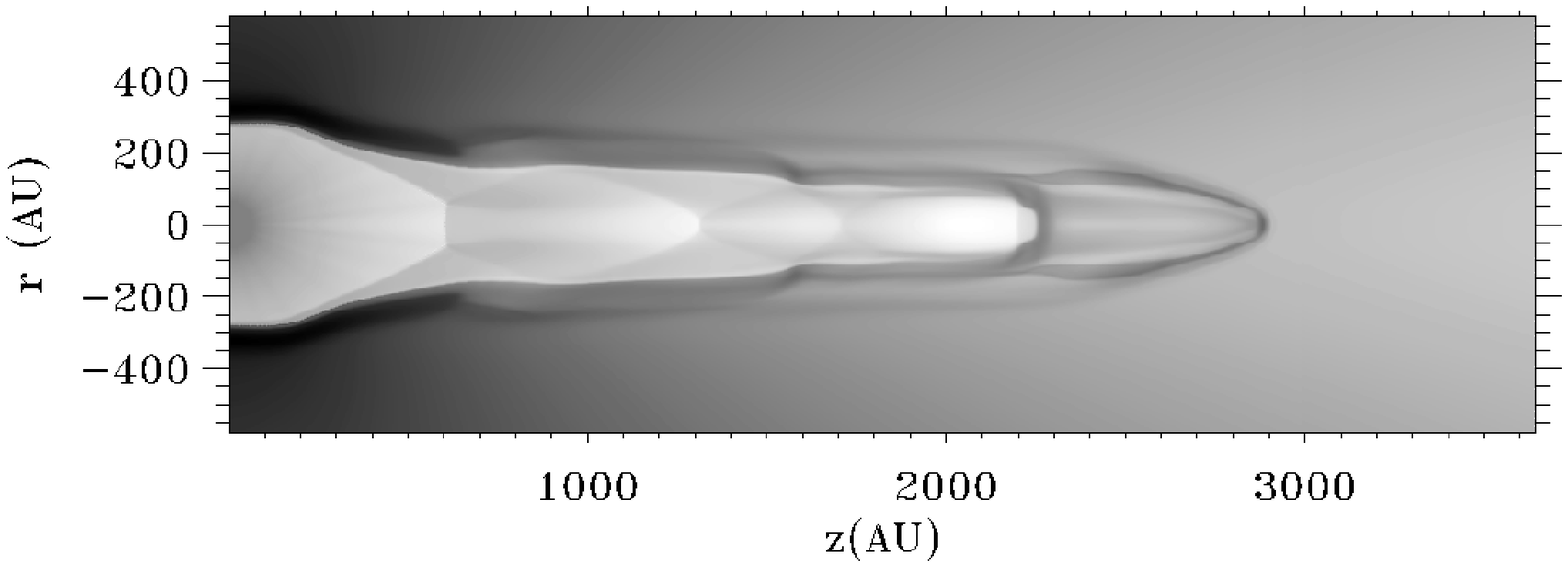,width=17cm}}
\centerline{\psfig{figure=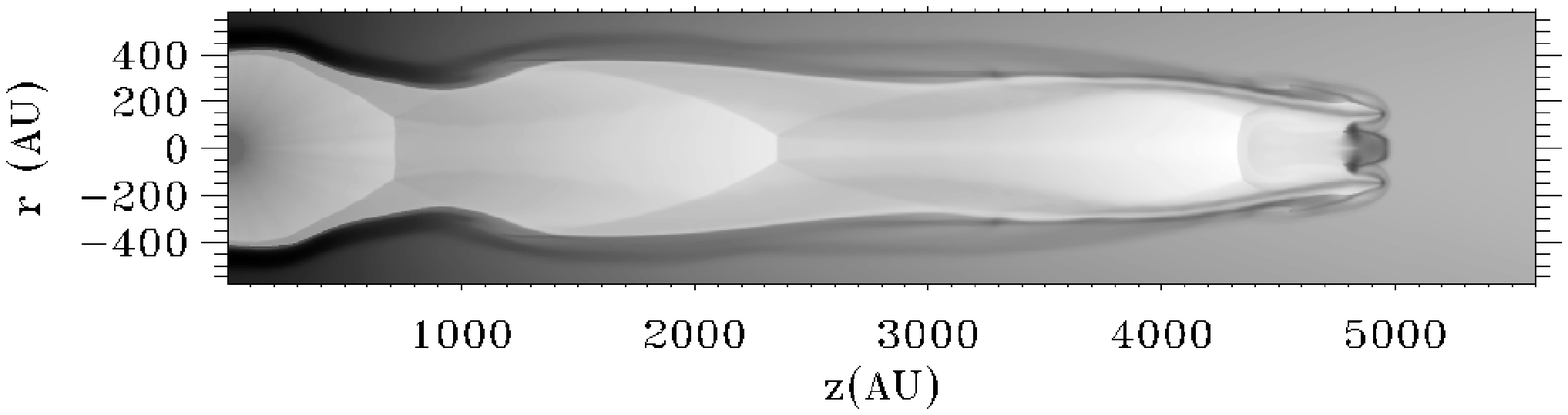,width=17cm}}
\caption{Density evolution for run A. The grey scales
 show $\log_{\rm 10}$ contours for times $t=$25, 62, 87, 137,
 225~years. The darker shades are higher density. One sees the initial
 radiative phase, the development of the `cool jet' and the
 development of the `hot jet'. The minimum/maximum pairs in units of
 cm$^{-3}$ are 
 $(152,4.95\times 10^{6})$, 
 $(80.2,4.05\times 10^{6})$, 
 $(76.3,3.78\times 10^{6})$, 
 $(24.8,3.20\times 10^{6})$, 
 $(4.38,1.976\times 10^{6})$. 
 See also Fig.~2.}  
 \label{runAdens}
\end{figure*}

\subsection{Cool jet} 
The evolution of the flow pattern in run~A is shown in Figs.~1 and 2. They
show the density and temperature in logarithmic grey scales. One sees how
initially the interaction creates an aspherical bubble, which is almost
completely radiative. The freely expanding wind extends almost all the way to
the shell of swept up ambient material. In the temperature plot (Fig.~2) one
can see the thin cooling layer separating the outflow and the swept-up
material and a small reservoir of hot gas at the very tip of the bubble.

As time progresses, a unique feature forms at the top of the bubble (Figs.~1b
and 2b). This structure takes the form of a dense, high velocity jet, with a
relatively low temperature, ($T \approx 10^4$ K), which is still higher than
that of the surroundings. By tracking the advection of a passive fluid tracer
we are able to identify the location of wind and ambient fluids at all
times. This shows that the jet is composed only of wind material. The
formation of this `cool jet' is seen in all the TVD-code and the Roe-solver
simulations where the cooling time scales are short.  In addition the `cool'
jet formation occurs for a large range of outflow velocities as we will show
below.  The collimation process we see in our simulations appears to be
related to the jet formation mechanism originally suggested by Cant{\'o} \&
Rodr{\'\i}guez~\shortcite{CanRod80} and later studied by Cant{\'o}, Tenorio-Tagle \&\
R{\'o}\.zyczka~\shortcite{CanTenRoz88} (hereafter: CTTR) and Tenorio-Tagle et
al.~\shortcite{TenCanRoz88} (hereafter: TTCR) using idealized analytical and
numerical models. In the models of Cant{\'o} and collaborators a jet is formed by
flows converging at the top of a radiative wind-driven bubble. A shock forms
at the apex of the converging flow which redirects material into a narrow
beam. We note that in the studies of CTTR and TTCR the converging beams were
imposed as initial conditions. In our results they are a natural and robust
consequence of the interaction between the outflow and the surrounding
material.

\begin{figure*}
\centerline{\psfig{figure=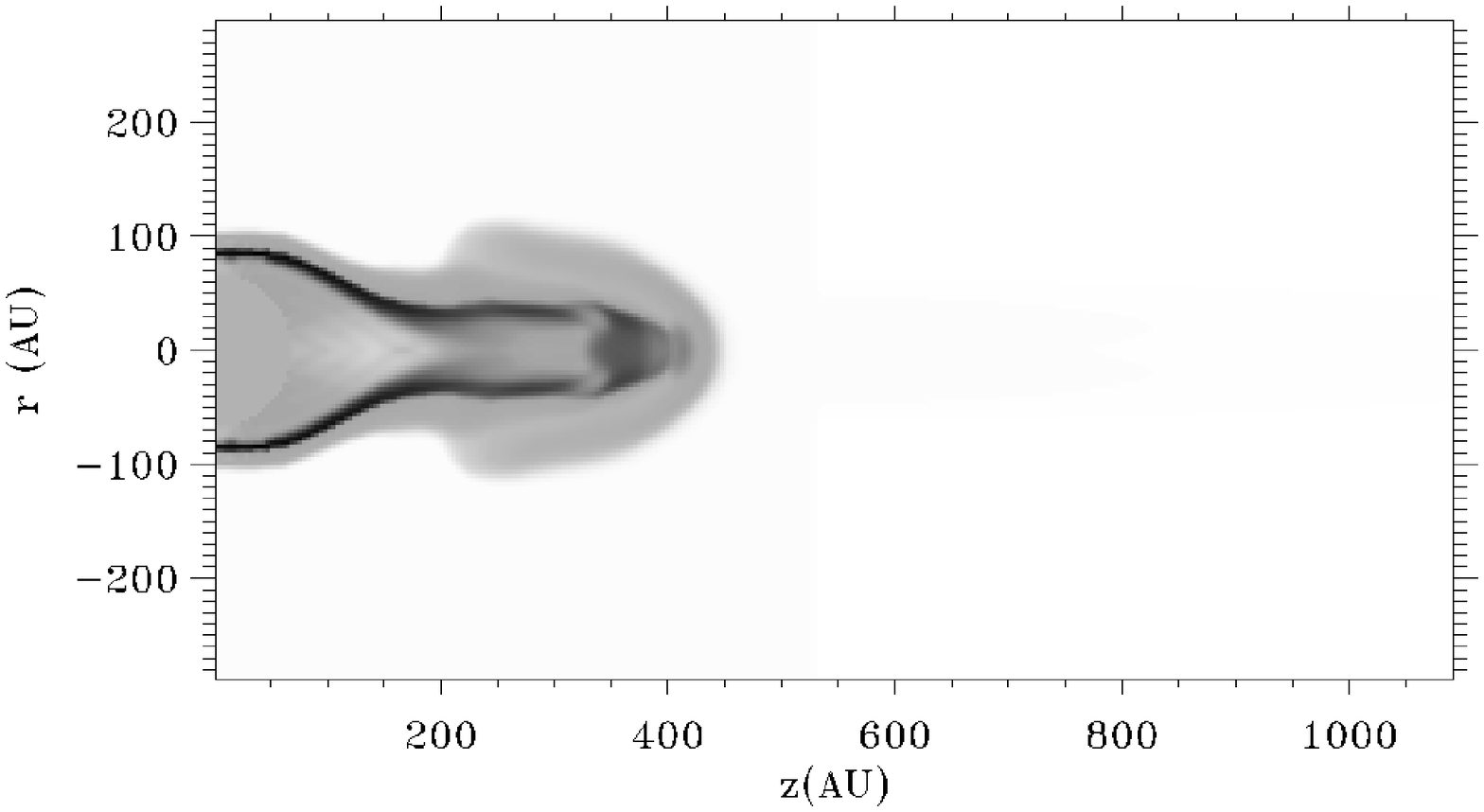,width=9cm}
\psfig{figure=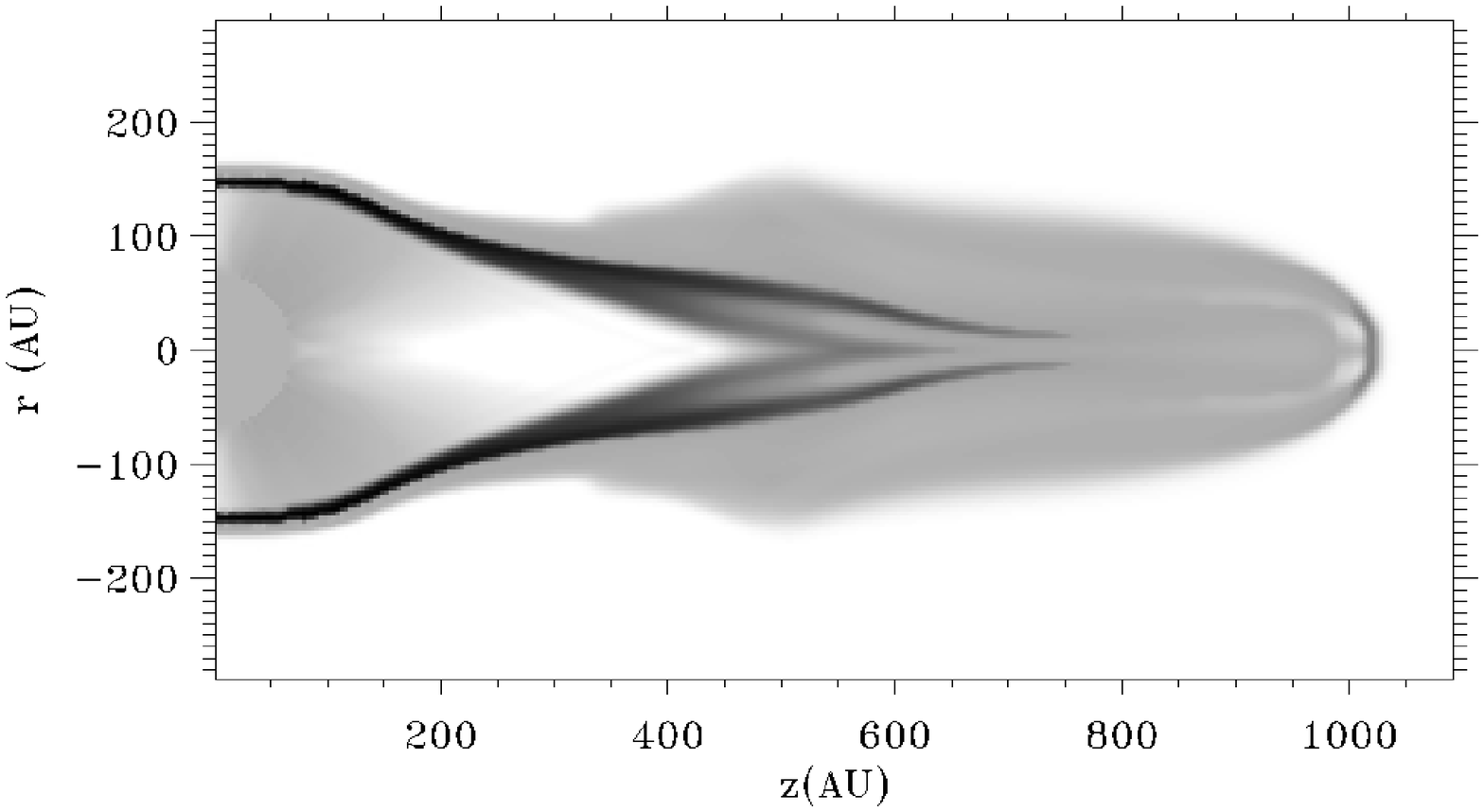,width=9cm}}
\centerline{\psfig{figure=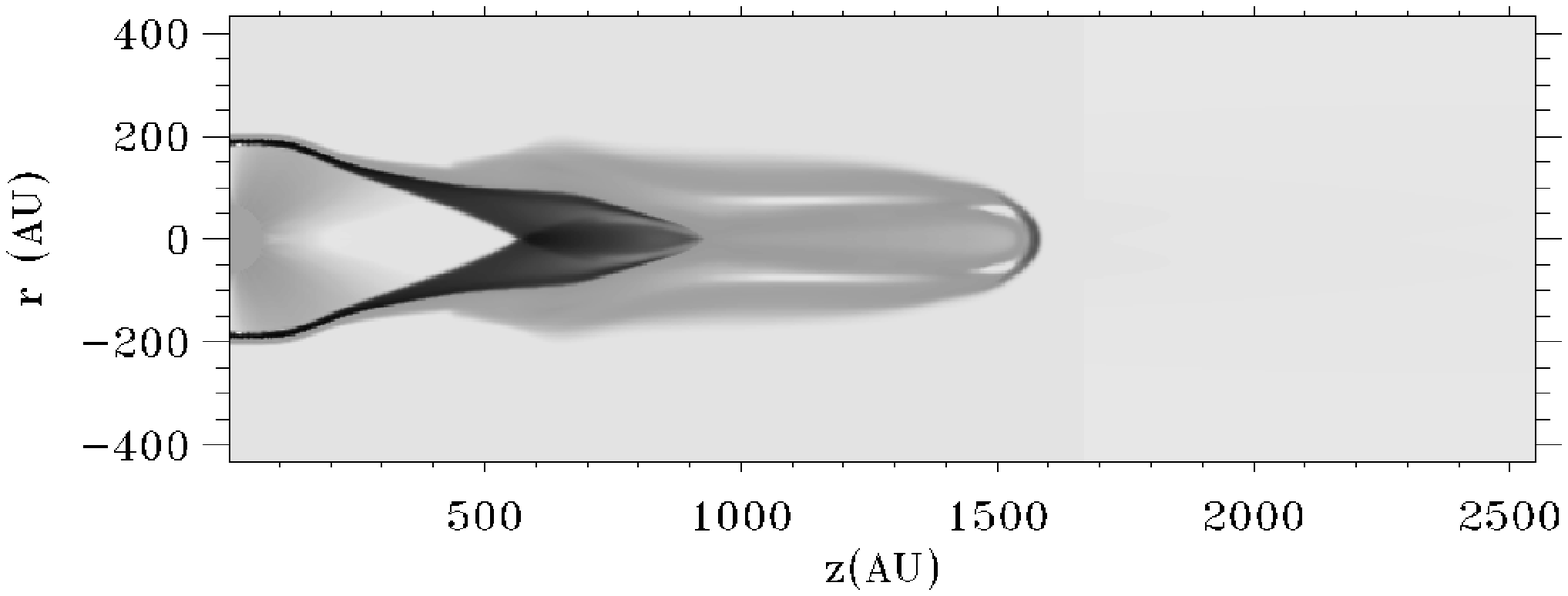,width=17cm}}
\centerline{\psfig{figure=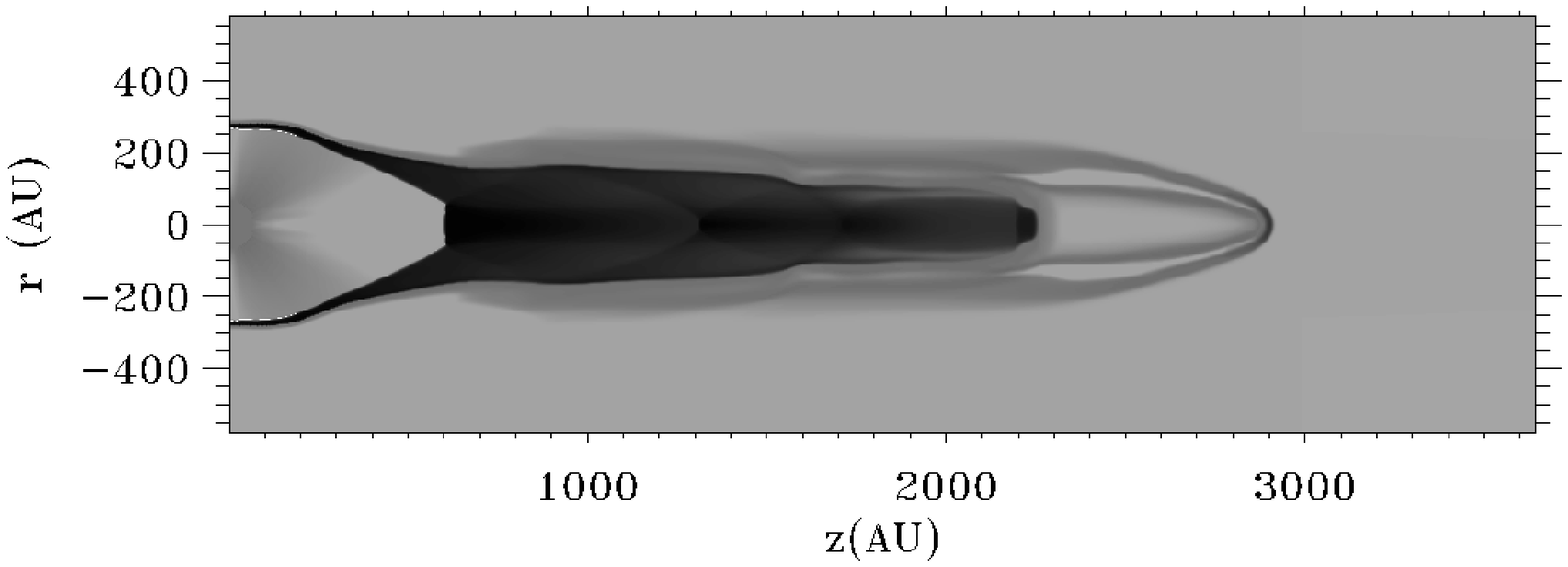,width=17cm}}
\centerline{\psfig{figure=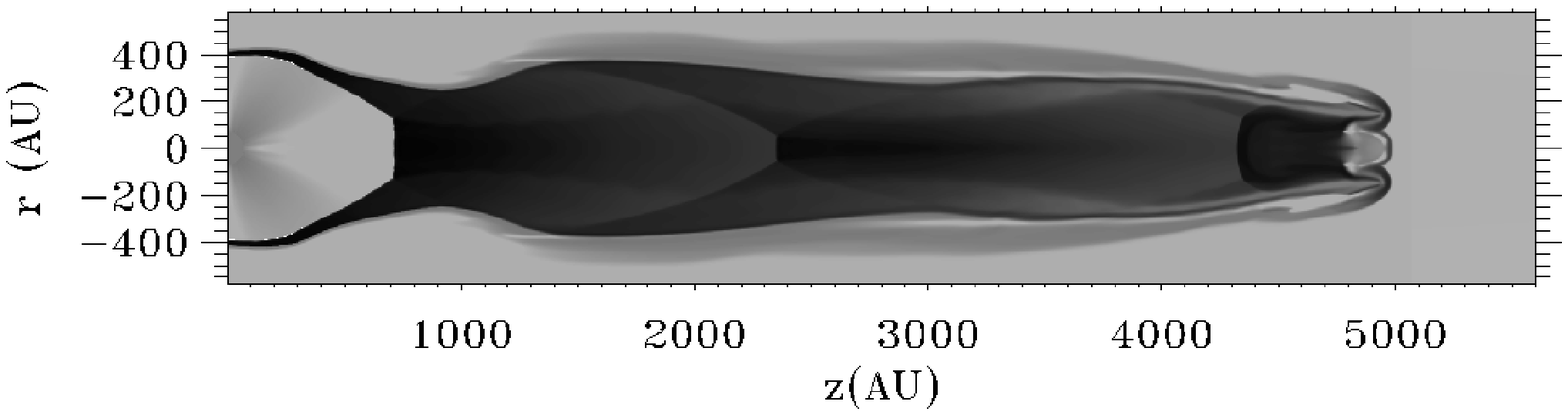,width=17cm}}
  \caption{Temperature evolution for run A. The grey scales
  show $\log_{\rm 10}$ contours for times $t=$25, 62, 87, 137, 225~years.
  The darker shades are higher temperatures. Note the presence of high
  temperature gas filling much of the volume in the first frame. The
  minimum/maximum pairs in units of K are $(993, 1.98\times 10^{6})$,
  $(988, 1.86\times 10^{6})$, $(429, 2.76\times 10^{6})$,
  $(12.4,2.97\times 10^{6})$, $(28.6,  3.12\times 10^6)$.} 
  \label{runAtemp}
\end{figure*}

The angle of incidence (between the beam and the symmetry axis) is quite low
here.  This can be seen in Fig.~3, which shows a contour plot of the
direction of the flow at the base of the jet ($t=62$~years). The quantity
plotted is the direction of the flow vectors, the contours running from
$-6^\circ$ to $0^\circ$, and negative angles implying a direction towards the
symmetry axis.  One can see how there is a beam flowing towards the axis at
an angle of about $-5^\circ$, and how a jet (with flow angle close to
$0^\circ$) forms at the axis. An angle of incidence of $5^\circ$ is lower
than any of the cases explored by TTCR But low angles are actually a good
condition for the mechanism, since not much kinetic energy is lost in the
shock. In fact the shock is so weak here that it is hardly noticeable.  At
the same time the collimation is very effective, something that could already
be seen in the results of CTTR and TTCR.

Although the same idea lies behind the work of CTTR and TTCR and the results
presented here, there are some important differences. In CTTR the bubble was
assumed to be in pressure equilibrium with the surroundings, which is not
the case here. This has implications for the long term evolution of these
structures, something we will come back to in Section~4. Both the analytical
models of CTTR and the numerical models of TTCR were further simplified by
assuming a steady state configuration at the base of the flow with
completely homogeneous beams colliding at the symmetry axis. Our flow is
time-dependent and far from homogeneous.  The converging flows which appear in
our simulations have density, velocity and pressure varying with
position. The idealization we do have in common with the previous work is
the assumption of a perfect, `rigid' symmetry axis.

\begin{figure}
  \psfig{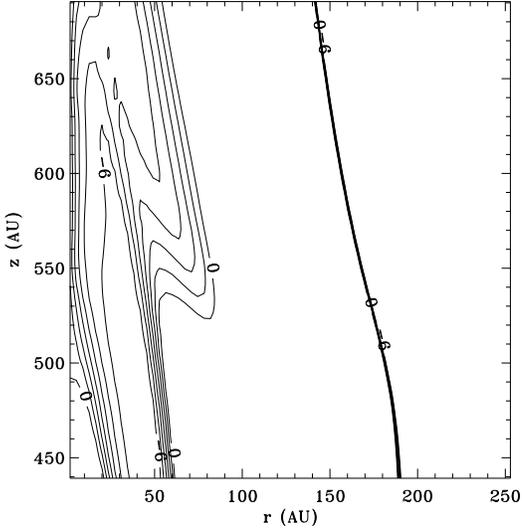}
  \caption{Contour plot of the angle of flow for run
  A. Only the area with the converging flow is shown. The symmetry
  axis is on the left. Contours are $-6^\circ$, $-5^\circ$, $-4^\circ$,
  $-3^\circ$, $-2^\circ$, $-1^\circ$, $0^\circ$. An angle of $0^\circ$
  means flow parallel to the symmetry axis and negative values mean flow
  pointing towards the symmetry axis.}
 \label{runAcanto}
\end{figure}

Despite these differences it is still instructive to make a simple comparison
between the analytic models and the behaviour in the simulations. The input
parameters for the CTTR model are the initial density $\rho_0$, width $y_0$,
and angle of incidence $\theta$ of the converging flow as well as the inverse
compression ratio $\zeta$ across shock at the tip of flow ($\zeta = 1/4$ for
a non-radiative shock). CTTR derived the following equations for the jet
radius $r_{\rm j}$, length $d_{\rm j}$, density $\rho_{\rm j}$, and velocity
$v_{\rm j}$:
\begin{equation}
r_{\rm j} = {y_0 \tan {\alpha} \over \tan\alpha+\tan\theta}\,,
\end{equation}
\begin{equation}
d_{\rm j} = {y_0 \over \tan\alpha+\tan\theta}\,,
\end{equation}
\begin{equation}
\rho_{\rm j} = \rho_0 {1 + \tan\theta/\tan\alpha \over \zeta}\,,
\end{equation}
\begin{equation}
v_{\rm j} = v_0 \cos(\theta + \alpha)/\cos\alpha\,,
\label{cttr1}
\end{equation}
where $\alpha$ is the half opening angle of the shock at the tip of the
converging flow or in other words the angle between the shock and the
symmetry axis,
\begin{equation}
\tan\alpha = {(1-\zeta)-\sqrt{(1-\zeta)^2-4\zeta\tan^2\theta}
\over 2\tan\theta}\,.
\label{cttr2}
\end{equation}
Taking values from the situation shown in Fig.~3 we have an angle of
incidence $\theta$ of around $5^\circ$ and a beam velocity $v_0$ of
approximately 300~\kms. For a qualitative comparison we assume the adiabatic
case (i.e.\ using an inverse compression factor of $\zeta = 0.25$ since the
shocks are weak) which leads to a jet opening angle $\alpha$ of $1\fdg 7$, or
nearly perfect collimation. The width of the beam $y_0$ at the convergence
point ($z = 540$ AU) is measured to be approximately 12 cell sizes (or $6.5
\times 10^{14}$ cm), leading to a jet cross section $r_{\rm j}$ of 3 cell
sizes (or $1.6 \times 10^{14}$ cm) and a length $d_{\rm j}$ of 102 cell
sizes (or $5.6 \times 10^{15}$ cm), both approximately consistent with the
result in the simulation. For the derived values of $\theta$ and $\alpha$ the
jet velocity $v_{\rm j}$ turns out to be 0.99 of the beam velocity, so
$\sim 300$~km~s$^{-1}$, and the shock velocity $v_{\rm s}^0$ a factor 0.12
of that, so around 36~km~s$^{-1}$, making it approximately a (weak) Mach 2
shock. The density of the beam is far from homogeneous, ranging from 1000 to
5000~cm$^{-3}$ which, still following the CTTR model, would lead to jet
densities of 16,000 to 80,000~cm$^{-3}$.  In our simulation the jet density
varies from 20,000 to 50,000~cm$^{-3}$. So, within all the uncertainties the
match between the analytic description and the simulation is quite good,
showing that it is indeed the convergence of conical flows that produces the
initial jet in these simulations.

Following the evolution of this `cool jet' we find that the bow shock of the
jet has a velocity of $\approx$ 110~\kms.  The flow velocity in the jet beam
lies between 160 and 200~\kms\ and typical values for the Mach number are
around 16 (at $t=62$~years), and 35 (at $t=112$~years), so the jet is highly
supersonic. The temperature in the main body of the jet is around 1,000~K
(but several 100,000 at the head), and the density is typically 2,000~\cc, at
$t=112$~years (though ten times as large in the earlier phases, $t\sim
60$~years). The width of the main channel is about 100~AU (five to ten times
the width of the initial focusing region), the wings adding another 50~AU.

As we said in Section~2, resolution effects may affect the details of the
numerical results. The presence of density gradients in the focusing region
may be real or may be due to grid-mixing. The same is true for the role of
instabilities (particularly Kelvin-Helmholtz) which may not be captured in
these simulations. Future studies will need to address these points using
higher resolution studies.

\subsection{Hot jet} 

The further evolution of the jet is influenced by another change in the
structure of the flow pattern. At about $t=80$~years a high temperature
($T\sim 10^6$~K) region develops at the top of the bubble and the base of the
jet (Figs.~1c and 2c). The emergence of this high temperature gas marks the
transition from a radiative to an adiabatic configuration.  This transition
starts at the poles because the densities are lowest there, and consequently
the cooling time longer. The transition from radiative to adiabatic can be
followed by looking at the shape of the wind shock. At $t = 62$ years it is
very aspherical with an an ellipticity of $e = 0.33$ where $e= R_{\rm
sw}(0^\circ)/R_{\rm sw}(90^\circ)$ and $R_{\rm sw}$ is the radius of the wind
shock.  As time progresses the shape relaxes. At $t=225$~years the wind shock
has reached an ellipticity of only $e = 0.5$. We note however that FM96 found
some degree of asphericity even in the fully non-radiative case and that this
was sufficient to produce collimation.

The high pressure region at the top of the bubble develops its own jet
structure in much the same way as was described in FM96. Although the gas is
decelerated at the wind shock, the constriction in the flow channel (the
contact discontinuity) quickly re-accelerates the flow so that it almost
reaches the original wind velocity (350~\kms).  A DeLaval nozzle is not
needed however. The obliqueness of the shock relative to the free streaming
wind at lower latitudes is high enough that post-shock material is focused
towards the axis but never becomes subsonic. This focusing effect for
aspherical wind shocks was described in FM96 where we predicted that the
collimated flow {\it behind} the oblique shock could retain its supersonic
character.  The calculations presented in that paper also demonstrated that
this effect becomes stronger when some degree of post-shock cooling is
included. This is exactly what is observed in the simulations presented
here. This also shows that the inclusion of cooling does not invalidate the
results obtained for the non-cooling case.  In fact, as was argued in FM96,
cooling enhances some of hydrodynamic collimation effects, such as the
asphericity of the wind shock.

This second jet, or `hot jet', has a Mach number at the base of about 3 to 4.
It catches up with the slower `cool jet' and in the last frame has almost
completely overtaken it ($t=225$~years). In the mean time it develops
similarly to the hot jets described in Paper~I, forming internal working
surfaces. These working surfaces are not stationary, but travel along the jet
with pattern velocities of around 50 to 100~\kms, not unlike the observed
ones.  As the jet grows, material in its beam continues to cool and the
temperature drops from about $2\times 10^6$~K just after an internal working
surface to about $2\times 10^5$~K just before the next one.  Since the
velocity does not change the Mach number of the jet grows as material
traverses the beam reaching values as high as $M \approx 8$ just before an
internal working surface. A vector plot of the velocity field at
$t=225$~years is shown in Fig.~4.

\begin{figure*}
 \psfig{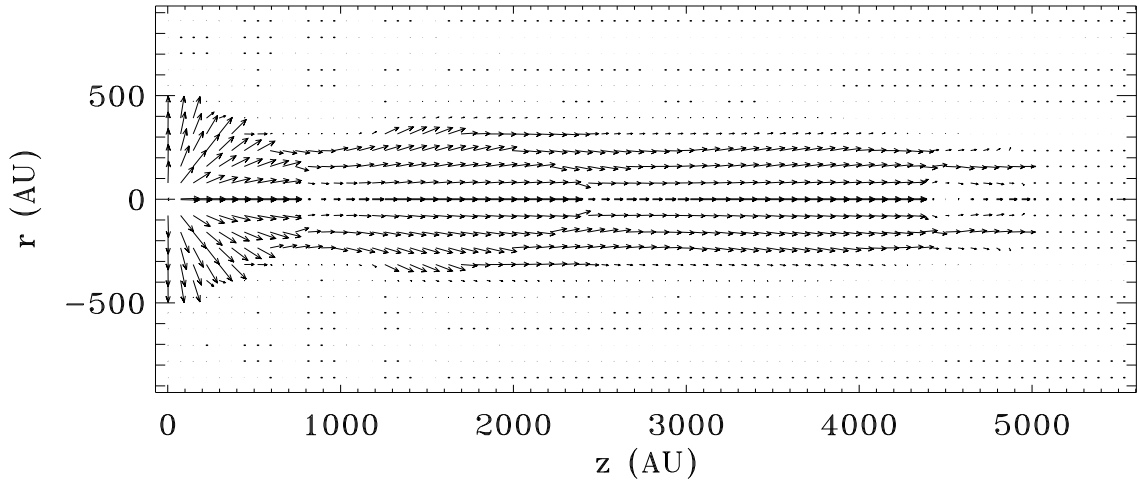} \caption{The velocity
 field for run A at $t=225$~years.  The length of the vectors is
 proportional to the absolute value of the velocity. The maximum value
 is 350~\kms\ and the minimum 0~\kms. One can see the focusing at the
 base of the jet and near the internal working surface.}
 \label{runAvel}
\end{figure*}

Thus, our simulation shows a sequence of two jets, an early `cool jet'
formed by the focusing of the outflow along the inner shock of a
momentum-driven bubble, and a later `hot jet' formed by the SFIC mechanism
described in FM96. As the system evolves this second jet overtakes the
first.

\subsection{Different wind velocities}

Run B has almost the same parameters as run A except for a mass loss rate
thats higher by a factor of 2 and a wind velocity which is 250~\kms. Because
of this lower wind speed one expects cooling to play a more important role
($T_{\rm shock} \propto V_{\rm shock}^2$) and we do find that in this case
the cool jet phase persists for a far longer time (about a factor of 5). We
note however that the same sequence of cool to hot jet that was seen in run A
also occurs in B and the results of runs A and B look similar when they have
reached similar size scales.  Run C with a 450~\kms\ outflow also develops
both a cool and a hot jet phase but here the cool jet phase lasts only for a
short time. Fig.~5 shows the comparison of the jets for runs B and C.
 
\begin{figure*}
\centerline{\psfig{figure=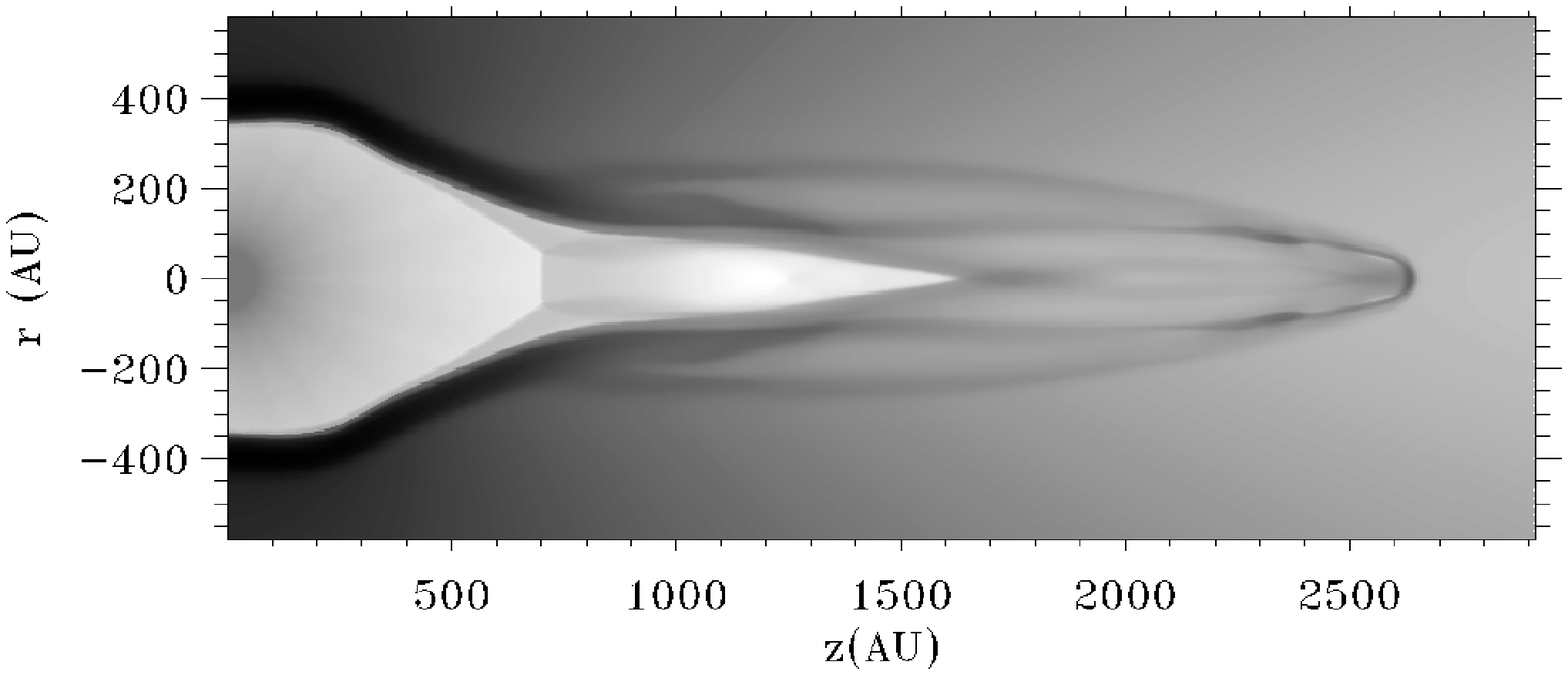,width=17cm}}
\centerline{\psfig{figure=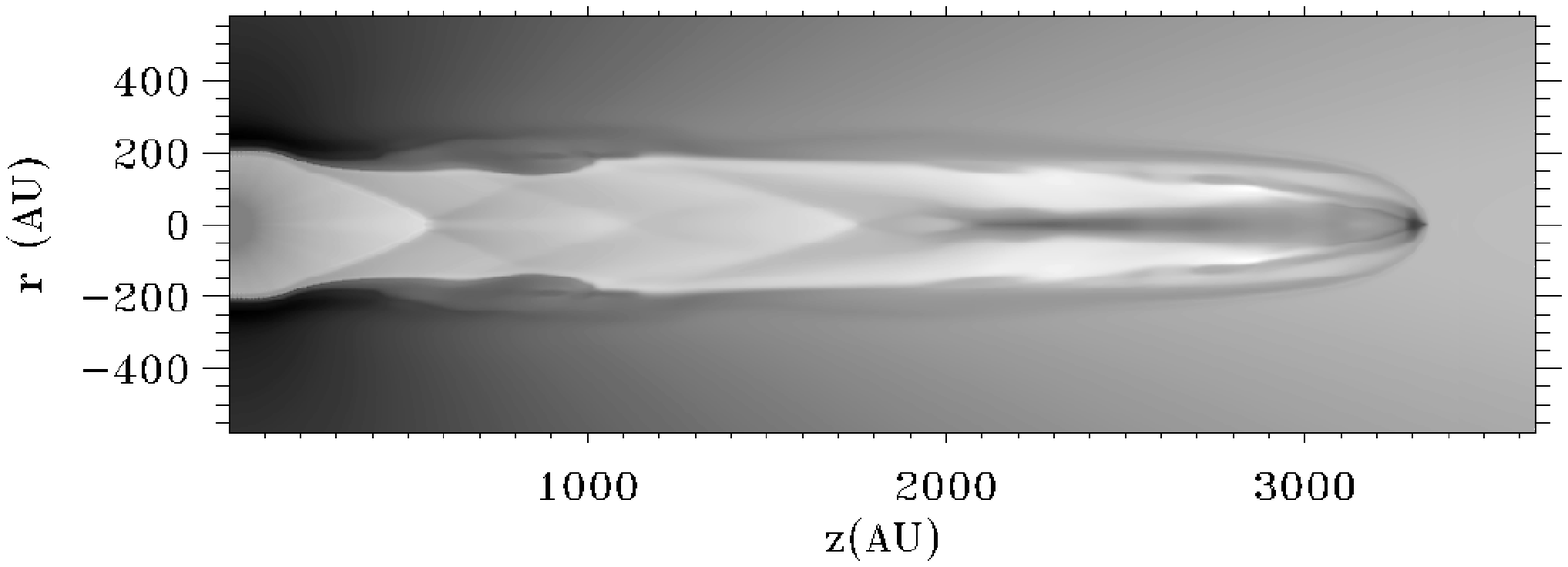,width=17cm}}
 \caption{Comparison between runs B and C. The grey scales
  show $\log_{\rm 10}$ density contours for run B ($t=210$~years) and
  run C ($t=80$~years). In run C the `cool jet' has been overtaken by
  the `hot jet' and the whole structure is much narrower.  The
 minimum/maximum pairs in units of cm$^{-3}$ are $(39.0,
 2.11\times 10^{6})$ and $(12.0,3.90\times 10^6)$.}
 \label{runBC}
\end{figure*}

Obviously the velocity is an important parameter for setting the time scales
of the different phases, but at the same time these results show that if
there is substantial cooling a cool jet will form, even at outflow velocities
as high as 450~\kms. In fact for these high velocities the jet stays
narrower; see Fig.~5.

\subsection{The importance of cooling}

To illustrate the importance of cooling for our collimation mechanism we
performed simulations with the similar parameters, but differing a factor 100
in absolute density in both the outflow and the surroundings (runs D and
E). The difference in density implies a factor $10^4$ less cooling in run
E. The simulation was done on a grid of cylindrical spherical coordinates,
using the Roe solver, and serves also a check that a different method
produces similar results. In Fig.~6 we show the result for the two runs at
the same time ($t=76$~years). Note first the lower value of the equator to
pole density contrast, $q=50$, in these runs compared with the previous
simulations.  In run D one sees the same structure as described above, a
momentum driven bubble with a jet forming at the tip. The jet is again very
narrow, in fact hardly resolved. In run E we see a very different
situation. The lack of cooling has caused the formation of an extensive
bubble of hot shocked wind material. The inner shock is still somewhat
aspherical, and a moderate degree of collimation occurs, but it is clearly of
a different character than in run D. This is the type of collimation that was
described by FM96. The comparison of these two runs shows that cooling is
essential in creating these type of collimated outflows.

The results from run D also show that the inner shock region is sensitive to
an instability in which ripples form along it. This behaviour is more
noticeable in this simulation mainly because of the less diffusive character
of the Roe solver method. The instability does not appear to have a large
impact on the overall flow pattern. Run E and the simulations in FM96 display
similar instabilities which lead to the formation of a hollow column of dense
material, surrounding the jet beam. This collimating `chimney' is not seen in
runs A--D. This may be due to cooling making the walls stiffer to the
instabilities or it may simply be a resolution effect. We do not believe we
have adequate grid cells to resolve the post-shock cooling region which has
$d_{\rm cool} < 10^{13}$~cm \cite{BlFrKo90} to the extent that the fate of
instabilities can be accurately tracked.  Regardless of the fate of the
chimney in the context of these simulations the phenomena may be important in
jet producing environments where radiative cooling is expected to be
unimportant such as in active galactic nuclei.

\begin{figure*}
\centerline{\psfig{figure=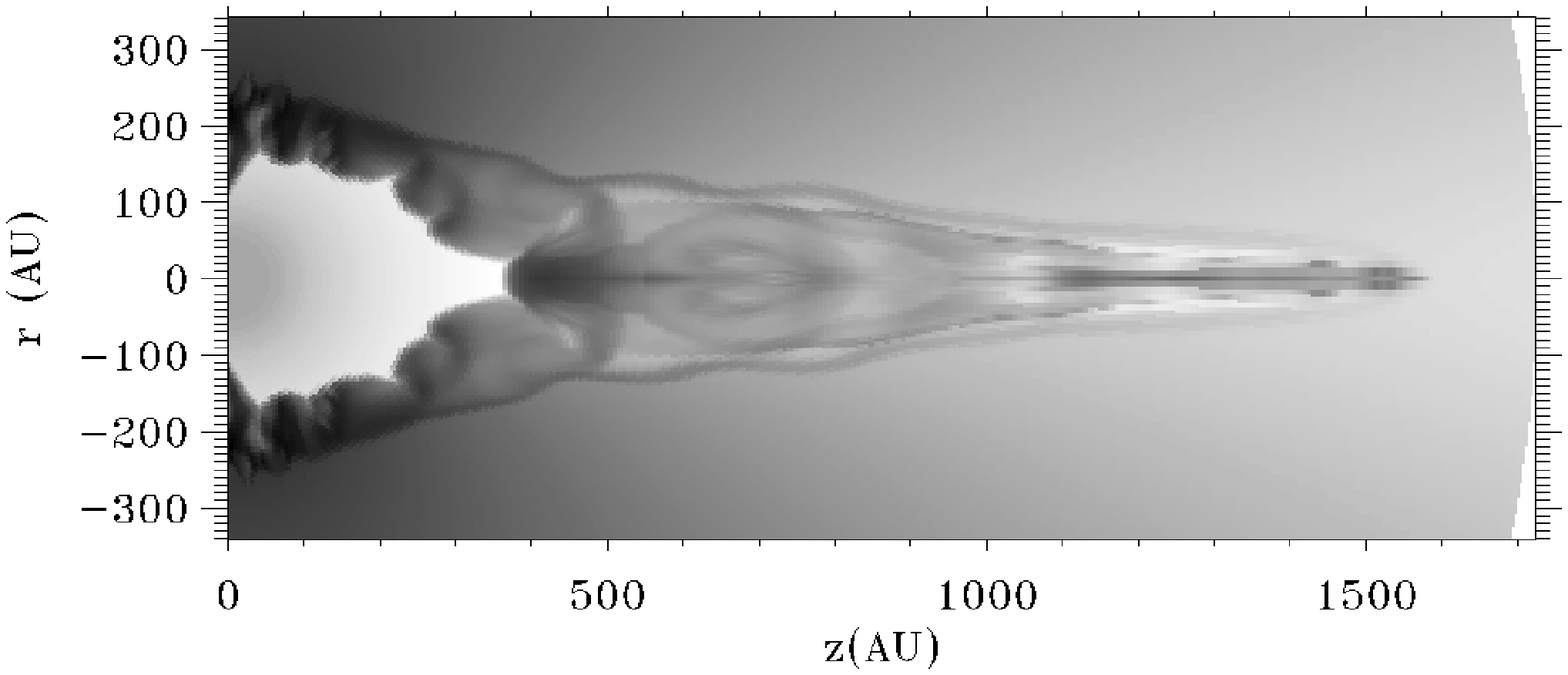,width=14cm}}
\centerline{\psfig{figure=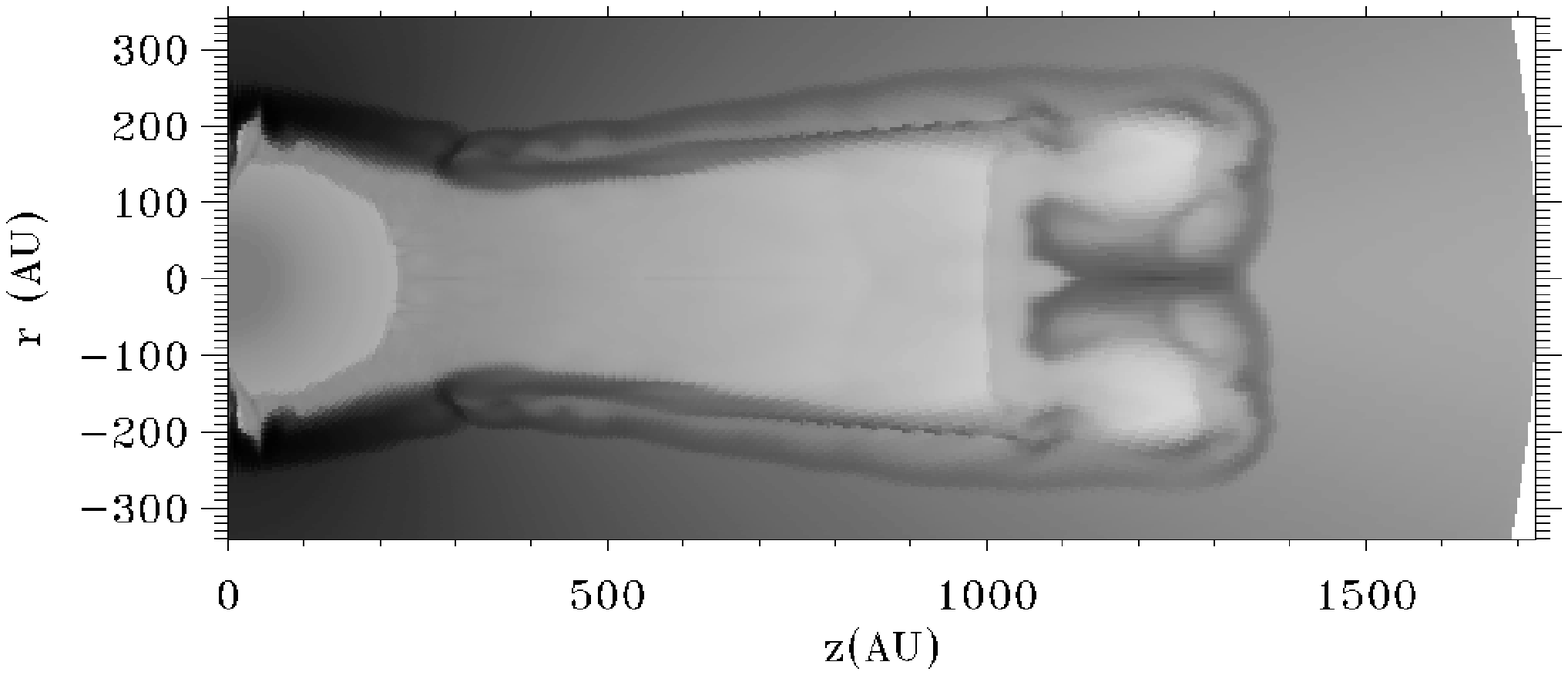,width=14cm}}
 \caption{Comparison between runs D and E. The grey scales
  show $\log_{\rm 10}$ density contours for run D and run
  E (both at $t=76$~years). Run D is still completely radiative, and a
  cool jet has formed, while in run E an extensive hot bubble has
  developed.   
  The minimum/maximum pairs in units of cm$^{-3}$ are $(412, 1.01\times
  10^6)$ and $(0.195, 6.65\times 10^4)$}.
 \label{runDE}
\end{figure*}

\subsection{Jet temperature}

Although the cooling time-scale for the material in the hot jet beam is
longer than the time it takes to traverse the jet (in these simulations),
the cooling time-scale is still relatively short. For run B, $t_{\rm cool} <
500$ years at the base of flow.  For run C, $t_{\rm cool} < 50$ years at the
base of flow.  Since observed jet size scales are much larger than our
computational grid ($L \approx 1$ pc) the material in the beam may be able
to cool to lower temperatures on time-scales shorter than the jet lifetime.
This tendency is apparent in run C where the cooling length $d_{\rm cool} =
v_{\rm j} t_{\rm cool} < 5 \times 10^{16}$~cm. Fig.~5 clearly shows the
presence of dense gas in the beam. This gas has cooled after having passed
through the wind shock.  We note also that we have used wind mass loss rates
that are on the lower end of the observationally accepted spectrum
\cite{Natta89,Ciccarea97}.  Higher mass loss rates mean high
densities, shorter cooling time scales which for our models imply longer
periods of cool jet collimation.

The issue of cooling the jets is an important point as they bear directly on
the observational consequences of our model.  The emission properties of a $T
= 10^6$ K plasma are obviously quite different than those for gas with $T <
10^4$ K.  For outflows from low mass stars current observations favour a cool
jet scenario (Hartigan, private communication). There is however clear
evidence for emission from ionized gas in the form of radio jets
\cite{Rodr95}.  For high mass stars the situation is more complicated.
The high velocities expected from their stellar winds almost ensures that
any shocks in the wind will drive the gas to very high temperatures ($T >
10^6$ K).  Thus the thermal state of the gas is a critical diagnostic for
our models.  In future papers we plan to address this issue by examining the
observational consequences of the various models using a `synthetic
observations' approach (cf.~Frank \& Mellema 1994). At the same time more
observational work on the temperatures of the jets and outflows is needed.

%
\section{The Effect of Wind Variability} 
The results from the previous section demonstrate that strong collimation
can be achieved from purely hydrodynamic interactions between winds and
protostellar environments.  But to apply this model to young stars we must
account for the time scales inherent to YSO jets and molecular outflows,
assuming that jets are connected to the outflows \cite{Chernea94}.  Recent
deep exposure images of HH jets such as HH34 \cite{BalLev94} and HH46/47
\cite{Heathea96} reveal multiple bow shock structures implying jet
lifetimes of many thousands of years or more. In addition, the molecular
outflows have dynamical lifetimes on the order of $10^3$ to $10^5$ years
\cite{Bach96}

If our simulations were allowed to run for such long periods we would expect
the equatorial regions of the wind blown bubble to expand until
the confining medium is eventually swept away.  In the process the jet
radius $r_{\rm j}$, which is of the order of the bubble's equatorial radius
$R_{\rm eq}$ would grow to an unacceptably large size.  Thus if our model is
to be viable it must account for the observed long lifetimes, small cross
sections, and small collimating length scales of the jets. In this section
we provide arguments that our model can account for these scales in a way
which recovers additional aspects of YSO jets as well.

First we note that our simulations are simplified in that they do not
include the presence of an accretion disk (of radial extend $\approx 100$
AU). In a realistic model it is unlikely that a dense but thin disk,
offering a relatively small cross section and high inertia to the wind,
would be pushed away.  In addition for low mass stars it is likely that the
wind itself may be originating in the inner regions of the disk
\cite{Shuea94}.  Above and beyond the disk the bubble can be constrained
in a number of ways.  Magnetic fields threading the circum-protostellar
environment can provide additional pressure restraining the growth of the
bubble.  If the field lines are anchored to a disk then magnetic tension can
be quite effective in constraining the expansion of the bubble
\cite{KwTad95}.

A more attractive and potentially more realistic means for stopping the
equatorial growth of the bubble comes directly from the observations.  As
was mentioned above many HH jets show clear signs of variations in velocity
along the jet beams.  In the most extreme case the jet appears to be
temporarily shut off, which explains the multiple bow-shock structures.  In
the HH34 superjet Bally \& Levine~\shortcite{BalLev94} find a periodicity in the beam of
$\tau_{\rm j} = 900/V_{\rm j300}$ years where $V_{\rm j300}$ is the jet
velocity in units of 300~\kms.  Thus it is likely that $300 < \tau_{\rm j} <
900$ years.  In addition smaller scale variations with periods $\tau < 100$
years are seen in many HH jet beams (Morse, private communication).  It
is reasonable that the velocity variations in the jet beam are a measure of
velocity variations in the source material of the jet. If this is the case
then one should consider a scenario where the the wind-blown bubble is
driven by a time-variable wind.  In our simulations we did not include the
effect of either gravity or the inward directed accretion flow.  Both of
these effects will decelerate the flow and constrain the bubble,
particularly near the equator where the bubble radius is small so that the
accretion velocity ($\propto R^{-0.5}$) and gravitational force density is
high.

\begin{figure}
  \psfig{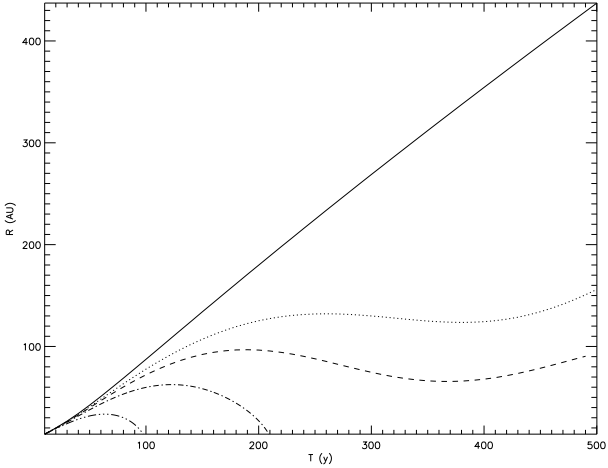}
 \caption{Evolution of spherical wind blown bubbles in accretion
  environment.  Shown are the radius of 5 bubbles driven winds with different
  periods as a function of time.  Solid line corresponds to a period of $P =
  10^6$ years. Dotted line: $P = 500$ years. Dashed line: $P = 400$ years.
  Dashed-dot line: $P = 300$ years. Dashed-dot-dot line: $P = 200$ years.}
 \label{timevol}
\end{figure}

Thus it is possible that even though a bubble can be `launched' by a
protostellar wind the accretion ram pressure and gravitational deceleration
may be able to significant slow, halt or even reverse the expansion of the
bubble during a quiescent phase. To test this conjecture we developed a
simple model for the interaction of a periodic stellar wind with an
accreting environment. The model assumes spherical symmetry and strong
radiative losses from the wind and ambient shocks so that we can use a
thin shell approximation.  The mass of the shell $M_{\rm s}$ and its radius
$R_{\rm s}$.  The equations for mass and momentum conservation for a shell
of mass $M_{\rm s}$ and radius $R_{\rm s}$ are
\begin{equation} 
{d R_{\rm s} \over d t} = V_{\rm s}\,,\\ 
\label{ed1} 
\end{equation}
\begin{eqnarray} 
{d M_{\rm s} \over d t} =
  4 \pi {R_{\rm s}}^2 \rho_{\rm w}(R_{\rm s}) 
   (V_{\rm w} - V_{\rm s}) + \nonumber\\
  \quad\quad 4\pi {R_{\rm s}}^2 \rho_{\rm a}(R_{\rm s}) 
   (V_{\rm s} + V_{\rm a})\,,
\label{ed2} 
\end{eqnarray}
\begin{eqnarray} 
{d M_{\rm s} V_{\rm s} \over d t} =
  4\pi {R_{\rm s}}^2 \rho_{\rm w}(R_{\rm s}) V_{\rm w} 
    (V_{\rm w} - V_{\rm s}) - \nonumber\\
  \quad\quad 4\pi {R_{\rm s}}^2 \rho_{\rm a}(R_{\rm s}) V_{\rm a}
    (V_{\rm s} + V_{\rm a}) - 
    {G M_* M_{\rm s} \over {R_{\rm s}}^2}\,,
\label{ed3} 
\end{eqnarray} 
\cite{VoKw85}. Using the following definitions we can rewrite equations
\ref{ed2} and \ref{ed3} in the form of a simple set of coupled
ordinary differential equations (ODEs)
\begin{equation} 
\rho_{\rm w}(R_{\rm s}) = {{\dot M_{\rm w}} \over 4 \pi {R_{\rm s}}^2
V_{\rm w}}\,, 
\label{eqdef1} 
\end{equation} 
\begin{equation} 
\rho_{\rm a}(R_{\rm s}) =
{{\dot M_{\rm a}} \over 4 \pi {R_{\rm s}}^2 V_{\rm a}(R_{\rm s})}\,, 
\label{eqdef2} 
\end{equation}
\begin{equation} V_{\rm a}(R) = \sqrt{{2G M_*  \over {R_{\rm s}}}}\,,
\label{eqdef3}
\end{equation} 
\begin{equation} 
{d M_{\rm s} \over d t} = {{\dot M_{\rm w}} \over V_{\rm w}}(V_{\rm w} -
V_{\rm s}) + {{\dot M_{\rm a}} \over V_{\rm a}} (V_{\rm s} + V_{\rm a})\,,
\label{ed5} 
\end{equation} 
\begin{eqnarray} 
{d V_{\rm s} \over d t} = {1 \over M_{\rm s}} \left[
{\dot M_{\rm w} \over V_{\rm w}} (V_{\rm w} - V_{\rm s})^2 - 
{\dot M_{\rm a} \over V_{\rm a}} (V_{\rm s} + V_{\rm a})^2 - 
\right. \nonumber\\
\quad\quad \left.{G M_* M_{\rm s} \over {R_{\rm s}}^2}\right]\,.
\label{ed6} 
\end{eqnarray}
Equations \ref{ed1}, \ref{ed5} and \ref{ed6} together with initial
conditions and a prescription for $V_{\rm w} = V_{\rm w}(t)$ define our
model.  For the wind velocity we take
\begin{equation}
V_{\rm w}(t) = V_{\rm wo}\left[{1 \over 2} \left(1 + \cos\left({2 \pi t \over \tau_{\rm w}}\right)\right) + .01\right]\,,
\label{vvardef} 
\end{equation}
and $\dot M_{\rm w}$ is kept constant. We also tried the case in which $\dot
M_{\rm w}$ varies in the same way as $V_{\rm w}$, which gives similar
results.

To solve our coupled set of ODEs we use a 4th order Runge-Kutta method with
an adaptive step-size.  The parameters for the wind ($\dot M_{\rm w}$ and
$V_{\rm wo}$) and the environment are the same as in Model A, see Table 1.
The initial radius at which the integration begins is $R_{\rm so} = 6.5
\times 10^{13}$ cm.  The initial mass of the shell $M_{\rm so}$ was taken to
be the amount of circumstellar mass originally contained in the volume $4
\pi R_{\rm so}^3/3$.  The initial velocity was arbitrarily taken to be
$V_{\rm so} = 5 V_{\rm a}(R_{\rm so})$.  The assumption is that before the
start of the integration, the bubble was set in motion, perhaps by an
energetic episodic outburst from the protostar. This initial condition gives
a total energy in the shell of $10^{40}$ ergs which is more than five orders
of magnitude less than what is released in a typical FU Orionis outburst
\cite{HartKen96}.

We calculated the evolution of the bubble for the case where the wind is
constant as well as for four time-dependent wind models with periods
$\tau_{\rm w} = $500, 400, 300 and 200~years.  The results are shown in
Fig.~7.  For the constant wind the bubble expands monotonically although it
does experience changes in velocity due to accretion ram pressure and
gravitational forces.  When the wind is allowed to vary these forces produce
dramatic changes in the bubble's evolution.  For all four variation periods
we see that the expansion of bubble can be reversed ($V_{\rm s} < 0$) for
some time.  For longer periods the bubble gains enough momentum before the
wind enters a minimum to either continue a slow expansion ($\tau_{\rm w} =
500$ years) or maintain a constant average radius ($\tau_{\rm w} = 400$
years).  For shorter periods the bubble is ``crushed'' during the wind's
quiescent phase by the inward directed forces. Note that we end the
calculation if the shock radius $R_{\rm s}$ became smaller than
$10^{12}$~cm.

Two important points should be kept in mind.  Firstly, the results are
sensitive to the initial conditions.  Whether a bubble expands, oscillates
or collapses is a strong function of the initial position $R_{\rm so}$ and
velocity $V_{\rm so}$ of the shell, and the momentum input of the wind. But
for each set of $(R_{\rm so},V_{\rm so})$ it is always possible to find
parameters for the wind which will lead to one of these three solutions. The
second point concerns the extension of this model to non-spherical
bubbles. Our simulations show that the equatorial radius in an aspherical
bubbles is always less than that derived from a 1-D calculation (FM96).
Thus the results shown in Fig.~7 should be taken as an upper limit to the
equatorial size scale of the bubble and hence the scale of the collimation
region.

From these results we conclude that that, in principle, time varying winds
can produce wind-blown bubbles whose size never increases beyond some upper
limit.  If these bubbles produce jets through the hydrodynamic mechanisms
described in Section~3, the model age and collimation scales can be made
consistent with observations.  We imagine that during a periodic increase in
mechanical luminosity the YSO will begin driving a bubble which in turn
produces collimated jets.  As the wind speed varies the bubble either
oscillates around some average radius (producing variations in the jet beam)
or it collapses entirely. Jet production begins again when the momentum in
the wind has increased enough to produce another bubble.

These oscillating bubbles are likely to be subject to Rayleigh-Taylor
instabilities, especially around the turn-around time. How this
affects the bubble structure and jet formation is difficult to say on
the basis of the simple one-dimensional model. It is a potential
problem to this scenario which we intend to study in future papers.

%
\section{Conclusions}
The results presented in this paper show that the collimation of an outflow
from a central object through the interaction with a surrounding toroidal
density distribution can be very efficient. In FM96 we addressed the case of
non-cooling flows and showed that these can be collimated much more
efficiently than one would anticipate from a simple analysis. Here we
included the effects of cooling and found that some of the effects described
in FM96 still hold, but that there is an additional collimating effect
producing converging, collimated flows at the poles of the wind blown
bubble. This effect appears only to operate under cooling conditions,
presumably because only then is the distance between the wind shock and contact
discontinuity small enough to constrain shocked wind material into a narrow
converging flow focused towards the axis. The cooling also allows the wind
shock to take on more prolate geometries which enhances the flow focusing.

When cooling is effective the interaction between a spherical outflow and a
toroidal environment is found to consist of two phases. Initially the bubble
is nearly completely radiative and the focusing of the flow at the top of
the bubble creates a dense, cool jet. The basic mechanism for this is the
same one that was studied by CTTR although the circumstances
here are somewhat different from the ones described there. The collimation
achieved is very good, the jet is an order of magnitude narrower than the
radiative bubble. We call this this the `cool jet'.

This `cool jet' phase is followed by a `hot jet' which also forms at the top
of the bubble. In this second case the collimation mechanism is similar to
the one described in FM96. This second jet starts forming when the cooling
at the top of the bubble becomes less efficient due to lower values of the
density there. The hot jet follows the same path as the cool one and
eventually overtakes it. The collimating processes consist essentially 
of focusing at
the inner shock (including supersonic post-shock flow), and the inertia of
surrounding material.

The difference between the cool and hot jet phases may be particularly
relevant when comparing low mass and high mass star formation and their jets.
High mass stars are likely to possess high velocity winds; when these are
shocked cooling is likely to be ineffective at the high post-shock
temperatures.  The jets which formed this way may be fast and hot. The shocks
may also produce significant non-thermal radiation via the first-order Fermi
process \cite{Ip95,Henrikea91}.  Such non-thermal emission has already been
observed.  Reid et al.~\shortcite{Reidea95} found strong synchrotron emission
arising at the centre of a linear chain of maser sources. Their
interpretation is that the wind from a massive star is being redirected via
shocks into a jet which drives the maser sources. This closely matches what
we predict from our model. Recent high resolution images show the synchrotron
emission actually appears as a two sided jet emanating from the geometric
centre of the two maser flows \cite{Wilnerea97}.

We also showed that it is, in principle, possible to obtain long-lived jets
with the correct collimation scales if the wind is variable. In that case
the bubble can collapse back towards the star during the wind's quiescent
phase; new jets form during the next active outburst phase. This may
naturally explain the knots and multiple bow-shocks observed in some HH jets.

The current investigation completes our work on the fundamental
hydrodynamics of the time-dependent hydrodynamic collimation. The next step
will involve using more realistic proto-circumstellar environments and
including relevant diagnostics to make a comparison with observations.  This
latter point is crucial and must be carried out carefully.  The main
difference between our models and those which rely on collimation through
MHD effects is the presence of strong shocks.  In our our models oblique
shocks are essential for redirecting an uncollimated wind into a jet.  The
tracers of these shocks and the absorbing effect of the dense surrounding
medium must be calculated carefully if realistic comparisons between theory
and observations are to be made.

%
\section*{Acknowledgments}
We wish to thank Dongsu Ryu for making the TVD code available to us and his
generous help in adding cooling to the TVD code. We wish to thank Tom Jones,
Alex Raga, Lee Hartmann \& Mark Reid for the useful and enlightening
discussions on this topic.  Support for this work was provided by NASA grant
HS-01070.01-94A from the Space Telescope Science Institute, which is operated
by AURA Inc under NASA contract NASA-26555.  Additional support came from the
Minnesota Supercomputer Institute. GM acknowledges support from the Swedish
Natural Science Research Council (NFR).

%

\label{lastpage}

%
\end{document}